\documentclass[12pt]{iopart}
\usepackage[dvipdf]{graphicx}
\begin{document}
\def\be{\begin{equation}}
\def\ee{\end{equation}}
\def\bc{\begin{center}} 
\def\ec{\end{center}}
\def\bea{\begin{eqnarray}}
\def\eea{\end{eqnarray}}
\title[Loops of any size and Hamilton cycles in random scale-free networks]{Loops of any size and Hamilton cycles in random scale-free networks}

\author{Ginestra Bianconi\dag\ \ddag\ 
\footnote[3]{e.mail: gbiancon@ictp.trieste.it}
and Matteo Marsili\dag\  
\footnote[4]{e.mail: marsili@ictp.trieste.it}
}

\address{\dag\ The Abdus Salam International Center for Theoretical Physics, Strada Costiera 11, 34014 Trieste, Italy}

\address{\ddag\ INFM, UdR Trieste, via Beirut 2-4, 34014, Trieste,Italy }

\begin{abstract} 
Loops are subgraphs responsible for the multiplicity of paths going
  from one to another generic node in a given network.  In this paper
  we present an analytic approach for the evaluation of the average
  number of loops in random scale-free networks valid at fixed number
  of nodes $N$ and for any length $L$ of the loops.  We bring evidence
  that the most frequent loop size in a scale-free network of $N$
  nodes is of the order of $N$ like in random regular graphs while
  small loops are more frequent when the second moment of the degree distribution diverges. In particular, we find that finite
  loops of sizes larger than a critical one almost surely pass from
  any node, thus casting some doubts on the validity of the random
  tree approximation for the solution of lattice models on these
  graphs. Moreover we show that Hamiltonian
  cycles are rare in random scale-free networks and may fail to appear
  if the power-law exponent of the degree distribution is close to $2$
  even for minimal connectivity $k_{min}\geq 3$. 

\end{abstract}

%\submitto{\JSTAT}

% Comment out if separate title page not required
\maketitle

The scale-free network structure has been found in a number of social,
technological and biological networks as the skeleton of their
interaction \cite{RMP,Doro_book,Internet}.  The main property of
scale-free networks is to have a power-law degree distribution $P(k)
\sim k^{-\gamma}$ and second diverging moment, i.e. $\gamma\in(2,3]$.
To distinguish between different scale-free networks, recently, much
attention has been devoted to network motifs
\cite{UAlon,Milo,Laszlo_m}, i.e. subgraphs that recur with higher
frequency than in maximally random graphs with the same degree
distribution.  Among those, the most simple types of subgraphs are
loops \cite{Loops,Guido_cycles,Avraham}, i.e.  closed paths
of various length that visit each node only once.  Loops (or cycles)
are interesting because they account for the multiplicity of paths
between any two nodes. Therefore, they encode the redundant
information in the network structure.  Another discriminant aspect of
real scale-free networks is the presence of degree correlations
between linked nodes. Characteristic motifs in a graph and degree
correlations are in many real graphs not independent phenomena but
they depend on each other as it has been shown for small (up to
maximal connectivity) size subgraphs in
\cite{Loops_last,Dorogovtsev,Vazquez_m,VazquezBa}. Last but not least,
it has been observed that the distribution of loop sizes is intimately
connected with the thermal properties of lattice models defined on
that graph \cite{Parisi_book_stat_phys,Marinari}. On the other hand,
the analytic approach to these models relies on the assumption that
locally a random graph can be considered to have a tree like structure
\cite{Dorog_solution_Potts,Leone_etal_Ising,Ehrhardt_Marsili_JSTAT},
i.e. that loops of finite size are rare.

In this paper we present an analytic derivation of the average number
of loops of any size in a random scale-free network. Our motivation is
that the results on random networks provide a reference picture which
often captures key intuition which extends to correlated networks (see
e.g.  \cite{epidemics,epidemics_corr}). In addition, it provides
valuable information for the statistical mechanics of lattice model on
random graphs~ \cite{Parisi_book_stat_phys,Marinari}.

Let us first recall the classic results for regular random graphs,
i.e. random graphs with fixed connectivity of the nodes $k_i=c$ for
each node $i$.  A regular random graph contains a finite number of small
loops of size $L\ll \log(N)$, with average expected number 

\be 
{\cal
  N}_L=\frac{1}{2L}(c-1)^L \label{reg.small.eq} 
\ee 
and Poisson
fluctuations around the mean.  On the contrary for large loop sizes
$L\sim O(N)$ the number of loops goes as 

\be 
{\cal N}_L=\exp(N
\sigma(\ell))
\label{scaling.eq}
\ee
where $\ell=L/N$ and $\sigma(\ell)$ is a function having the maximum
at $\ell_{max}=c/(c+1)$ whose expression can be found in the
literature \cite{Janson,Marinari}. 
Regarding  Hamilton cycles, i.e. loops that span the entire
network $L=N$  their expected number for a large regular random graph
is diverging with the system size as long as $c\geq 3$. For $c=2$ the
average number of Hamilton cycles  goes to zero as the system size
diverges \cite{Janson}. 
Coming to the scale-free network literature, Ref.\cite{Avraham}
analyzes the number of loops of any size on a pseudo-fractal
scale-free graph and report the scaling behavior 

\be
\log {\cal N}_L=Lf(L/L^*)
\label{Nl}
\ee
with $L^*=N^{1/(\gamma-1)}$. No result were presented on Hamilton
cycles, to our knowledge, so far. 

In the following we characterize the statistics of loops in random
scale-free networks. We find a larger number of small loops with
respect to regular random graphs. In particular, we compute the
expected number of loops of a given size passing through a 
node and find that when $\gamma\in (2,3)$ this number diverges with
the network size, beyond a finite loop size. This raises some doubts
on the solution of lattice models on these graphs based on the local
tree approximation
\cite{Dorog_solution_Potts,Leone_etal_Ising,Ehrhardt_Marsili_JSTAT}.
We also find that loops have a 
 characteristic size $L^*\sim N$.  In other words, our results are consistent with the
scaling $(\ref{Nl})$ with $L^*\sim N$ and with Eq. (\ref{scaling.eq})
for regular graphs. This suggests that the result of Ref.
\cite{Avraham} crucially depends on the peculiar correlations of the
ensemble they consider. Special attention will be given to Hamilton
cycles that in networks with a small $\gamma$ exponent can fail to
exist unless the lower cutoff of the distribution is large enough.

There are different ensembles of random networks one can consider. The classical one follows the prescription of Molloy and Reed\cite{MR}: First, to each node $i$ of the network is assigned a connectivity $k_i$ drawn from the chosen probability distribution, secondly edges are randomly  matched.
This ensemble indeed generates networks of given degree distribution but it may yields networks with multiply occupied links. More precisely, the distribution of 
the links between two nodes of connectivity $k_i$, $k_j$ is a Poisson variable with mean $k_i k_j/(c N)$, where henceforth $c=\langle k \rangle$ will denote the average connectivity. Hence the probability of no multiply occupied links is
\be
\Pi_{i> j}\left(1+\frac{k_i k_j}{cN}\right)e^{-\frac{k_i k_j}{cN}}\simeq 
e^{-\frac{1}{4}\left(\frac{\gamma-2}{3-\gamma} m^{\gamma-2}K^{(3-\gamma)}\right)^2 }
\ee
where the right hand side refers to a scale free random graph with degree distribution $P(k)= a k^{-\gamma}$ with 
$k\in [m,K]$. Taking a structural cutoff $K\sim N^{1/2}$ \cite{poli1,FS_vesp}, we conclude that double links appear with probability one for $\gamma\in (2,3]$ as $N \rightarrow \infty$ in the  Molloy-Reed ensemble.
When counting loops of a network this effect becomes relevant and undesirable. Thus we consider also another ensemble where double links cannot appear: the static fitness network\cite{HV,HV2}.
In the fitness ensemble nodes are assigned a random variable (fitness)
$q$ drawn from a $\rho(q)$
distribution function and every couple of nodes is linked  with a probability depending on the fitness of the considered nodes $p(q,q')$.
When $\rho(q)$ is power-law distributed and $p(q,q')=\frac{q q'}{\langle q\rangle N}$ the resulting graph is  a random scale-free graph characterized by the same exponent of the fitness distribution. In these graph the connectivity of every node is a Poisson variable with expected value $\langle k(q)\rangle=q$.
This ensemble doesn't allow for networks with double links but instead
may give rise to networks with isolated nodes ($k_i=0$) or to nodes
connected with a single link ($k_i=1$) to the others. The presence
of such nodes rules out the possibility to find Hamilton cycles,
hence we shall take this effect into account when discussing Hamilton
cycles.
 
Consequently, the Molloy-Reed ensemble and the fitness ensemble are
not equivalent and have intrinsic properties that could perturb
sensitively the counting of the number of loops.  In order to
understand the dependence on the details of how graphs are generated
in the following we are going to study the expected number of loops in
the two ensembles.

\section{Loops in the fitness ensemble}
The prescription of Ref.\cite{HV} to generate a class of random scale-free networks with exponent $\gamma$ is the following: {\em i)} assign to each node $i$ of the graph a hidden continuous variable $q_i$ distributed according to a scale-free distribution
$\rho(q)=\rho_0 q^{-\gamma}$ for $q\in [m,Q]$ with $\rho_0=(\gamma-1)/(m^{1-\gamma}-Q^{1-\gamma})$ the normalizing constant. 
Then {\em ii)} each pair of nodes with hidden variables $q,q'$ are linked with probability $qq'/(cN)$, where
$c=\langle q\rangle$ is the expected value of $q$. 
For $\gamma\leq 3$ the cutoff $Q\sim N^{1/2}$ is needed to keep the linking probability smaller than one, i.e. $Q^2/(cN)<1$ while for $\gamma>3$ the cutoff is the natural one $Q\sim N^{\frac{1}{\gamma-1}}$.
By construction the expected value of the connectivity of a node with hidden variable $q$ is $\langle k|q\rangle=q$ and there are no multiple connections between nodes. 
 Notice that the average connectivity of the graph is given by

\begin{equation}\label{avgk.eq}
  \langle k\rangle =\langle q\rangle = c\to \frac{\gamma-1}{\gamma-2} m
\end{equation}
in the limit $N\to\infty$.

A loop of size $L$ is an ordered set of distinct nodes $\{1_i,\ldots,i_L\}$. For each choice of the nodes, the
probability that they are connected in a loop is
\[
\frac{q_{i_1}q_{i_2}}{Nc}\frac{q_{i_2}q_{i_3}}{Nc}\cdots
\frac{q_{i_L}q_{i_1}}{Nc}=\prod_\ell \frac{q_{i_\ell}^2}{Nc}.
\]
The total number of possible loops joining these $L$ nodes, in any possible way is $L!/(2L)$ where the factor $2L$ comes from the fact that the initial node of the loop can be chosen in $L$ ways and that there are two orientations. In order to count loops, let us lump together nodes with hidden variable $q_i\in[q,q+\Delta q)$, where $\Delta q $ is a small interval of $q$. In each interval of $q$ there are $N_q\simeq NP(q)\Delta q$ nodes of the network. For each choice of the $L$ nodes, let $n_q$ be the number of nodes with $q_{i_\ell}\in [q,q+\Delta q)$. Then the average number of loops of size $L $ in the graph is given by the number of ways we can choose $\{n_q\}$ nodes multiplied by the probability that these nodes are connected in all distinguishable orderings. Consequently we have 
\be
{\cal N}_L=\frac{L!}{2 L} \sum_{\{n_q\}}  \prod_q \frac{N_q!}{n_q! (N_q-n_q)!}\left(\frac{q^2}{Nc}\right)^{n_q}
\label{uno.eq}
\ee
where the sum is extended over all $\{n_q\}$ such that $\sum_q n_q=L$. Introducing this constraint 
with a delta function and using its integral representation, we find
\be
{\cal N}_L=\frac{L!}{2 L}
\int_{-\infty}^\infty \frac{dx}{2\pi} e^{iLx+N\left\langle \log\left[1+q^2e^{-ix}/(Nc)\right]\right\rangle}.
\label{exact.eq}
\ee 
Notice that in Eq. (\ref{exact.eq}) one can safely take the limit $\Delta q\rightarrow 0$ and that the average over the $P(q)$ distribution is taken assuming that we focus on the limit $N\to\infty$. In what follows, we will evaluate Eq. (\ref{exact.eq}) in different ranges of $L\le N$ in the limit $N\to\infty$.
 
\subsection{Small loops}
For $L$ finite but large, the integral in Eq. (\ref{exact.eq}) is dominated by values $x\simeq -iz^*$ where
\begin{equation}\label{apprx.small.eq}
  e^{-z^*}\simeq\frac{\langle q^2\rangle}{Lc}\left(1-\frac{\langle q^4\rangle}{\langle q^2\rangle^2}\frac{L}{N}+\ldots\right)
\end{equation}
where we have neglected all terms beyond the first leading correction when $N\to\infty$. The argument of the exponential in Eq. (\ref{exact.eq}) can be expanded around $x\sim -iz^*$ yielding 
$N\left\langle \log\left[1+q^2e^{-ix}/cN\right]\right\rangle+Lix\simeq 
L\left[1-z^*-\frac{1}{2}(x-iz^*)^2+O(x-iz^*)^3\right]$.
Hence the integral can be estimated by saddle point for $L$ large. Using the asymptotic expression $L!\simeq\sqrt{2\pi L}L^L e^{-L}$, we find to leading order

\begin{equation}\label{Nsmall.eq}
  {\cal N}_L\simeq \frac{1}{2L}\left(\frac{\langle q^2\rangle}{c}\right)^L
\end{equation}
This approximation is valid as long as the leading correction in Eq. (\ref{apprx.small.eq}) is small.
Using that $\langle q^n\rangle=\rho_0(Q^{n-\gamma+1}-m^{n-\gamma+1})/(n-\gamma+1)$ for $\gamma\ne n$ and that 
$Q\sim \min(N^{1/2},N^{\frac{1}{\gamma-1}})$ we find that the expression above for ${\cal N}_L$ holds when 
\begin{equation}\label{small.diseq}
  L\ll N\frac{\langle q^2\rangle^2}{\langle q^4\rangle}\sim\left\{\begin{array}{cc}
    N & \gamma>5\\
    N^{\frac{\gamma-3}{\gamma-1}} & \gamma>3 \\
    N^{(3-\gamma)/2} & 2<\gamma<3 \
  \end{array}\right.
\end{equation}
with logarithmic corrections for $\gamma=3$ and $5$.
Note that strictly speaking the expansion $(\ref{apprx.small.eq})$ is converging only for  $N\langle q^2\rangle/L \gg N$, i.e. $L\ll N^{(3-\gamma)/2}  $ for $2<\gamma<3$ and $L\ll N^{(\gamma-3)/(\gamma-1)}$  for $\gamma>3$.
Nevertheless Eq.$(\ref{Nsmall.eq})$ remains valid in the limits $(\ref{small.diseq})$ as an asymptotic expansion. 
 For $\gamma>3$ we obtain a result very similar to Eq. (\ref{reg.small.eq}) for regular graphs. On the contrary, for 
$2<\gamma<3$ we have $\langle q^2\rangle\simeq a N^{(3-\gamma)/2}$, with $a$ a constant, 
hence the number of finite loops 
\begin{equation}\label{Nsmall.sf.eq}
  {\cal N}_L\simeq \frac{1}{2L}\left(\frac{a}{c}\right)^L N^{\frac{3-\gamma}{2}L}
\end{equation}
diverges as $N\to\infty$.

\subsection{Intermediate loop sizes and the most frequent loops} 

It is convenient, at this point, to write Eq. (\ref{exact.eq}) as

\begin{equation}\label{f.eq}
  {\cal N}_L\simeq\int_{-\infty}^\infty \frac{dx}{2\sqrt{\pi L}}e^{Nf(Nce^{ix},L/N)}
\end{equation}
where we have used Stirling's approximation and

\begin{equation}\label{fyl.eq}
  f(y,\ell)=\left\langle \log\left[1+q^2/y\right]\right\rangle+\ell\log (\ell y/c)-\ell.
\end{equation}

The integral can be computed by saddle point method, deforming the contour of integration so as to pass from the point
where $f$ is stationary. The condition $\partial_yf(y,\ell)=0$ yields 
\be
\left\langle\frac{q^2}{q^2+y}\right\rangle=\ell.
\ee
Let $y^*(\ell)$ be the value of $y$ which solves this equation. We can expand $f(Nce^{ix},\ell)$ around the
corresponding (complex) value $x^*$ of $x$
\bea
f(Nce^{ix},\ell)&=&f(y^*,\ell)-\frac{{y^*}^2}{2}\frac{\partial^2 f}{\partial y^2}(x-x^*)^2+\ldots\\
&=&f(y^*,\ell)-\frac{y^*}{2}\left\langle \frac{q^2}{\left[q^2+y^*\right]^2}\right\rangle (x-x^*)^2+\ldots
\nonumber
\eea
As long as 
\[
Ny^*\left\langle \frac{q^2}{\left[q^2+y^*\right]^2}\right\rangle\gg 1
\]
we can neglect higher order terms. This yields the leading behavior

\begin{equation}\label{gen.eq}
  {\cal N}_L\simeq \frac{1}{2}
  \left[L Ny^*\left\langle \frac{q^2}{\left[q^2+y^*\right]^2}\right\rangle\right]^{-1/2}e^{Nf(y^*,\ell)}
\end{equation}

\begin{figure}
\begin{center}
%\epsfbox{width=75mm, height=55mm}{lmax.eps}
%\end{center}
\includegraphics[width=75mm, height=55mm]{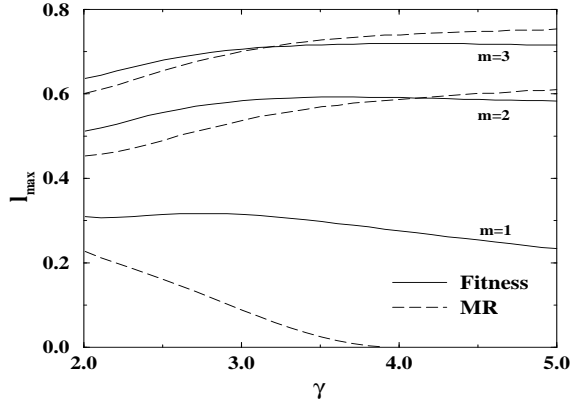}
\end{center}
\caption{\label{lmax.fig}The behavior of $\ell_{\max}$ as a function of $\gamma$ for  $m=1,2,3$ and $N=10^6$ for  fitness ensemble (solution of the Eq. $(\ref{lmax.eq})$ in solid lines) and for the Molloy-Reed ensemble (solution of the  Eq. $(\ref{lmax_MR.eq} )$ in dashed lines). }
\end{figure}

The number of loops ${\cal N}_L$ takes its maximum for loops of length $L=N\ell_{\max}$ where

\bea\label{lmax.eq}
  \left\langle\frac{q^2}{c+q^2\ell_{\max}}\right\rangle =1.
\eea
The solution $\ell_{\max}$ is plotted in Fig. \ref{lmax.fig} against $\gamma$ for scale-free graphs and different values of $m$, for $N=10^6$. Notice that as $\gamma\to 2^+$, the size of most probable loops vanishes as $\ell_{\max}\sim \gamma-2$.  Around the maximum, ${\cal N}_L$ takes a form similar to that for regular graphs (see Eq. \ref{scaling.eq}) which is consistent with the scaling form Eq. (\ref{Nl}) with $L^*\sim N$.

For scale-free random graphs with $2<\gamma<3$ there is an
intermediate range of loop sizes $L\sim N^{(3-\gamma)/2}$ which is
related to the solutions with $y^*=\mu N$ with $\mu> 1$. More
precisely, we find that for loops of size
$L=\chi(\mu)N^{(3-\gamma)/2}$ we have
\[
{\cal N}_L\simeq \frac{G(\mu)}{L} e^{L[\log(\mu L/c)-1+H(\mu)]}
\]
with $G(\mu)$ a function of $\mu$ and
\bea
\chi(\mu)=(\gamma-1)m^{\gamma-1}\sum_{n=0}^{\infty}(-1)^n\frac{\mu^{-n-1}}{3-\gamma+2n}\nonumber \\
H(\mu)=\frac{1}{\chi(\mu)}(\gamma-1)m^{\gamma-1}\sum_{n=1}^{\infty}\frac{(-1)^{n-1}\mu^{-n}}{n(1-\gamma+2n)}.
\eea
Notice that ${\cal N}_L$ does not satisfy a simple scaling form such as Eq. (\ref{Nl}) in this intermediate region.

\subsection{Hamilton cycles}

From Eq. $(\ref{uno.eq})$ we can  easily calculate  the number of  Hamilton cycles. Indeed for $L=N$ we have the asymptotic behavior
\be
{\cal N}_N=\sqrt{\frac{\pi}{2N}}  e^{N[2\langle\log q\rangle-1-\log c]}.
\label{h.eq}
\ee
This is the expected number of Hamiltonian cycles over all the
networks of the fitness ensemble including networks with nodes of low
degree $k_i=0,1$, which by definition cannot have an Hamilton cycle. 
It seems a sensible thing to compute the number of Hamilton cycles in
networks with a  minimal degree connectivity  grater than $3$, i.e. $k_i\ge=3$. 
In fact it  is well known that for a regular random graph of
connectivity $c=2$ the expected number of Hamilton cycles goes to zero
in the $N\rightarrow \infty$ limit whereas regular graphs are
Hamiltonian when $c\ge 3$. 
Taking this as a reference result, we 
normalize ${\cal N}_N$ by the probability $\pi$ that all the nodes
have at least 
$3$ connections. Since the connectivity of each node in the fitness network is aPoisson variable with expected value $q$ the probability that all the
nodes have connectivity $k\geq 3$ is simply given by
$\pi=e^{N\lambda(m,\gamma)}$, where 

\be
\lambda(m,\gamma)=\langle\log(1-(1+q+q^2/2)e^{-q})\rangle 
\ee
In the limit $N\to\infty$ we find 
\be
\frac{1}{N}\log\frac{{\cal N}_N}{\pi}\to
\log\left(\frac{\gamma-2}{\gamma-1}m\right)+\frac{3-\gamma}{\gamma-1}-\lambda(m,\gamma).
\label{hsf.eq}
\ee
This implies that if random scale-free graphs have Hamilton cycles
only for $m>m_c(\gamma)$ where $m_c(\gamma)$ is the value of $m$ for
which Eq. (\ref{hsf.eq}) vanishes. Conversely, for $m<m_c(\gamma)$ a
random scale-free graphs has almost surely no Hamilton cycle. 

In Fig. $\ref{Hc.fig}$ we report the critical value $m_c(\gamma)$ as a
function of $\gamma$.  Notice that $m_c\sim 1/(\gamma -2)\to\infty$ as
$\gamma\to 2^-$. Consequently, if we consider only the networks of the nesemble with $k_{min}\ge 3$, we find that random
graphs with $\gamma<\gamma_{\rm Fit}^*=2.16\dots$ (where
$m_c(\gamma_{\rm Fit}^*)=3$) do not have Hamiltonian cycles.
Considering that regular random graphs with $k=c\ge3$ are Hamiltonian,
this may seem a surprising result, at first sight. The basic intuition
to explain this apparent paradox is that most paths pass through well
connected nodes. Hence even if $k_i\ge 3$ it is very unlikely to
have a path spanning the entire network which is not passing through
the most connected nodes more than once.

\subsection{Loops passing through a node}

In order to  count of the number of loops of size $L$ passing through a given
node, with fitness value $q_i$, we can repeat the previous
outlined above, without taking the average over $q_i$. 
For $\gamma<3$  and short loops sizes $L\ll N^{\frac{3-\gamma}{2}}$
this gives the expected number 

\be
{\cal N}_L({q_i})\simeq \frac{{q_i}^2}{cN}\frac{1}{2L}\left(\frac{a}{c}\right)^{(L-1)}N^{\frac{3-\gamma}{2}(L-1)}.
\ee
Focusing on nodes with $q_i\simeq N^{\alpha}$, we find that the
number of loops of size 
\be
L\geq L_0\equiv 1+\frac{2}{3-\gamma}(1-2\alpha)
\ee
diverges with the network's size $N$. For example, nodes with a finite
$q_i$ have an infinite number of $L=5$ loops passing through it in
networks
with $\gamma<2.5$ but at most a finite number of loops of size $L=3$
if $\gamma>2$. The most connected nodes ($\alpha=1/2$) instead have an
infinite number of loops of any size $L\ge 3$ passing through it.
Notice that $L_0\to\infty$ as $\gamma\to 3$ in order to
match the behavior $L_0\sim \log N$ of regular graphs. Conversely, 
in a finite graph of $N$ nodes, only the large fitness nodes with 
\be
q_i\gg N^{\frac{1}{2}-\frac{3-\gamma}{2}(L-1)}
\ee 
belong to a significant number of loops of size $L$.

\section{Loops in the Molloy-Reed ensemble}

The counting of the number of loops in the Molloy-Reed\cite{MR}
follows a procedure much similar to the one considered for the fitness
ensemble nevertheless giving different results. 
To construct a Molloy-Reed network one proceed as follows:
{\it i)} a degree is assigned to each node of the network following
the desired degree distribution with cutoff $K\sim\min(N^{1/2},N^{\frac{1}{\gamma-1}})$. Degree distributions which do not
satisfy the parity of $cN=\sum_i k_i$ are disregarded; 
{\it ii)} the edges  coming out of the nodes are randomly matched
until all edges are connected. When this procedure ends with nodes
having links to themselves (tadpoles), the whole network is rejected
and the procedure is started anew.

To calculate ${\cal N}_L$ in this ensemble first one has to count in
how many ways it is possible to have a loop of size $L$ in the network
and weight the results with the fraction of possible networks in the
ensemble which  contains the loop. 
Let us first state that the total number of graphs in the Molloy-Reed
ensemble is given by $(cN-1)!!$. Indeed when constructing the network
by linking $cN$ unconnected edges one start by taking one edge at
random and connecting it to one of the $(cN-1)$ possible
connections. Then one proceed taking another edge and linking it to
one of the remaining  $(cN-3)$ possible connections thus giving rise
of one of the $(cN-1)!!$ possible networks. 
By similar arguments one shows that the total number of networks
containing a given loop of size $L$ are $(cN-2L-1)!!$. 
On the other side the total number of loops of size $L$ in the
Molloy-Reed ensemble are given by the number of ways one can choose an
ordered set of $L$ nodes  $\{1_i,\ldots,i_L\}$  of connectivity
$\{k_1,k_2,\dots,k_L\}$ and connect them on a loop. As for the fitness
network the total number of possible loops joining these $L$ nodes, in
any possible way is $L!/(2L)$. 
The number of ways one can choose the edges coming out of the nodes to form the loop is given by
\[
\Pi_{i=1}^L k_i(k_i-1).
\]
Consequently the average number of loops in the Molloy-Reed ensemble will be given by
\be
{\cal N}_L=\frac{L!}{2 L} \sum_{\{n_k\}}  \prod_{k=m}^K \frac{N_k!}{n_k! (N_k-n_k)!}\left({k(k-1)}\right)^{n_k}W_{N,L}
\label{unoMR.eq}
\ee
where $N_k=NP(k)$ ($n_k$) is the number of nodes with connectivity $k$
present in the network (loop), $K$ is the  cutoff of the degree distribution and  the sum over $\{n_k\}$ is restricted to  $\{n_k\}$
such that $\sum_k n_k=L$. Moreover we use the definition $W_{N,L}={(cN-2L-1)!!}/{(cN-1)!!}$.
If we use the Stirling approximation for  $W_{N,L}$ 
we get the expression
\be
 W_{N,L}\sim (cN)^{-L}  e^{Ng(\ell)}
\ee
with $\ell=L/N$ and
\be
g(\ell)=\frac{1}{2}(c-2\ell)\log\left(\frac{c-2\ell}{c}\right)+\ell.
\label{g.eq}
\ee
Thus we get
\be
{\cal N}_L=\frac{L!}{2 L} \sum_{\{n_k\}}  \prod_{k=m}^K \frac{N_k!}{n_k! (N_k-n_k)!}\left(\frac{{k(k-1)}}{cN}\right)^{n_k}e^{N g(\ell)}
\ee
which except for the substitution $q^2\rightarrow k(k-1)$ and the factor $\exp(Ng(\ell))$ is equivalent to the expression $(\ref{exact.eq})$  of the average number of loops of size $L$ in the fitness ensemble.
Following the same steps as in the fitness ensemble, we get 
\be
{\cal N}_L=\frac{L!}{2 L}
\int_{-\infty}^\infty \frac{dx}{2\pi} e^{iLx+N\left\langle \log\left[1+k(k-1)e^{-ix}/(Nc)\right]\right\rangle+N g(\ell)}
\label{exactMR.eq}
\ee 
with  $g(\ell)$ given by Eq. (\ref{g.eq}).

\subsection{Small loop size} The number of small loops in the Molloy-Reed ensemble is given by
\begin{equation}\label{NsmallMR.eq}
  {\cal N}_L\simeq \frac{1}{2L}\left(\frac{\langle k(k-1)\rangle}{c}\right)^L
\end{equation}
where this approximation is valid asymptotically for loops sizes satisfying Eq.$(\ref{small.diseq})$.
Note that as in the fitness ensemble for $\gamma\in (2,3)$ short loops diverge as ${\cal N}_L\sim N^{\frac{3-\gamma}{2}}$.
\subsection{Intermediate loops sizes} For intermediate loops  in the Molloy-Reed ensemble a similar expression to Eq. $(\ref{fyl.eq})$ holds 
with 
\be
f'(y,\ell)=\langle \log[1+k(k-1)/y]\rangle+\ell\log(\ell y/c)-\ell+g(\ell).
\ee 
The calculations of the  average number of loops is very similar for the Molloy-Reed and fitness ensemble, with a difference for the equation of the loops of maximal size which in the Molloy-Reed ensemble satisfy
\be\label{lmax_MR.eq}
   \left\langle\frac{k(k-1)}{c-2\ell_{max}+k(k-1)\ell_{\max}}\right\rangle =1.
\ee
In Fig. $\ref{lmax.fig}$ we report the value of $\ell_{max}$ in the Molloy-Reed
networks as a function of $\gamma$ for different value of the minimal connectivity $m$ for $N=10^6$.

\subsection{Hamiltonian cycles} 
\begin{figure}
\begin{center}
%\epsfbox{Hc.eps}
%\end{center}
\includegraphics[width=75mm, height=55mm]{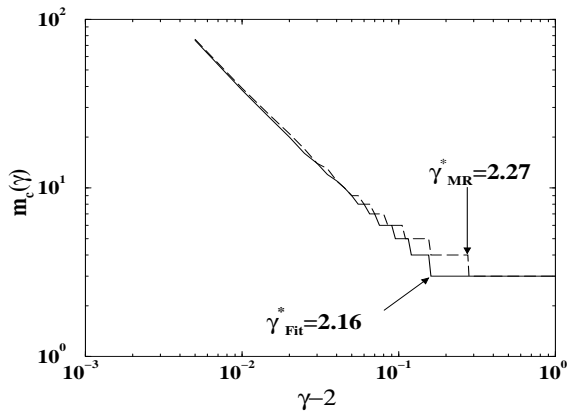}
\end{center}
\caption{\label{Hc.fig} 
Dependence of  $m_c(\gamma)$ on $\gamma$ in the fitness (solid line) and Molloy-Reed ensemble (dashed line) in the limit $N\rightarrow \infty$. Observe that for $\gamma<\gamma^*$  we have $m_c(\gamma)>3$ in both ensembles.}
\end{figure}

Starting with expression
 $(\ref{exactMR.eq})$ one can easily evaluate the expected number of
 Hamiltonian cycles in Molloy-Reed networks. 
 Indeed for $L=N$  and  $c>2$ one can use the Stirling approximation
 to find the asymptotic behavior ($N\rightarrow \infty$) 
\be
\frac{1}{N}\log({\cal N}_N)={\langle\log( k(k-1)/c)\rangle+\frac{1}{2}(c-2)\log(1-2/c)}.
\label{hcmr.eq}
\ee
If we approximate $k(k-1)$ with $k^2$ which is possible close to
 $\gamma\rightarrow 2$ in the limit $c\rightarrow \infty $ we recover
 the same behavior as in the fitness ensemble: if the minimal
 connectivities  $m$ is smaller than the value $m_c(\gamma)$ for which
 Eq. (\ref{hcmr.eq}) vanishes, then a scale-free network is typically not Hamiltonian.
 In Fig. $\ref{Hc.fig}$ we report $m_c(\gamma)$ for $2<\gamma<3$ and
 we confirm the behavior $m_c\sim 1/(\gamma-2)$ for $\gamma\to 2$. 
 For example, we find that Molloy-Reed random scale-free graphs with minimal connectivity
 $m=3$ are typically not Hamiltonian if $\gamma<\gamma^*_{\rm MR}=2.27\dots$.
As for the fitness ensemble, the intuition is that it is not possible
to extract from a random-scale-free graphs a
 subgraph which is a regular random graph with fixed connectivity
 $c\geq 3$ if $m<m_c(\gamma)$.

\subsection{Loops passing through a node} To count the loops of size $L$
passing through a given node of connectivity $k$ we must fix it and
choose other $L-1$ nodes to form the loop. 
For short loops sizes and exponent $\gamma<3$ this gives the expected number
\be
{\cal N}_L(k)\simeq \frac{k(k-1)}{cN}\frac{L-1}{L} {\cal N}_{L-1}
\ee
with ${\cal N}_L\sim N^{\frac{3-\gamma}{2}L}$. The same results
derived for the fitness ensemble hold: There is a critical finite loop size
$L_0(k)$ such that there are infinitely many loops of size $L>L_0$
passing through a given node of connectivity $k$. 
On the contrary, in a finite but large network of $N$ nodes, 
loops of size $L$ becomes significant for  nodes of connectivity
\be
k\gg N^{\frac{1}{2}-\frac{3-\gamma}{4}(L-1)}.
\ee

\section{Numerical results}

\begin{figure}
%\begin{center}
%\epsfbox{fit.eps}
%\end{center}
\begin{center}
\includegraphics[width=75mm, height=55mm]{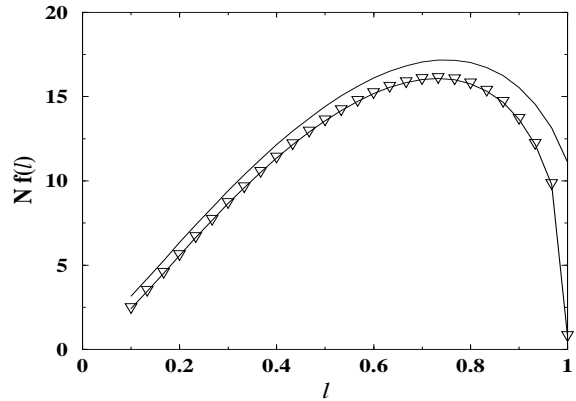}
\end{center}
\caption{\label{Fit.fig}Number of loops for fitness networks of
  $N=30$ nodes and given distribution  $\rho(q)$ with $m=3$. The
  average is taken over $50$ realizations.} 
\end{figure}

We compare the analytic results derived so far with the direct count
of the number of loops in a sample of computer generated random graphs
in both the fitness and the Molloy-Reed ensemble. This is important
because ${\cal N}_L$ is a fluctuating quantity which takes
exponentially large values. In other words, the analytic calculation
of the expected number of loops may be dominated by (exponentially in
$N$) rare realization of graphs with an exponentially large number of
loops. In this case the number of loops of a typical realization of a
graph would differ from our estimate.  We have chosen the fastest
known algorithm for calculating the total number of loops exactly
\cite{Johnson} as in Ref. \cite{Marinari}.  This algorithm has a upper
time bound of $O(N{\cal L})$ where ${\cal L}$ is the total number of
loops in the network.  The simulations performed in this way enable
one only to consider small networks sizes $N<50$ and small $m\le 3$ as
the total number of loops in such graphs increases exponentially with
the system size.  Note that for such small sizes the degree
distribution contains nodes of very similar degree since the upper
cutoff is $K\sim6 $ for $\gamma=2.1$.  Moreover in order to compare
the direct counting with the analytical calculation, we have chosen a
fixed degree (fitness) distribution $N_k=NP(k)$ to reduce fluctuations
that become relevant for such small sizes.  In Fig.  $\ref{Fit.fig}$ we report the
analytic prediction of the average number of loops of a given size in
a fitness network of $N=30$ nodes.  This results are compared with
direct counting of the loops in computer realizations of these
networks were data are averaged over $50$ realizations. We found
strong sample to sample fluctuations which we believe are responsible
for the deviation from the analytical results.

\begin{figure}
\begin{center}

\includegraphics[width=75mm, height=55mm]{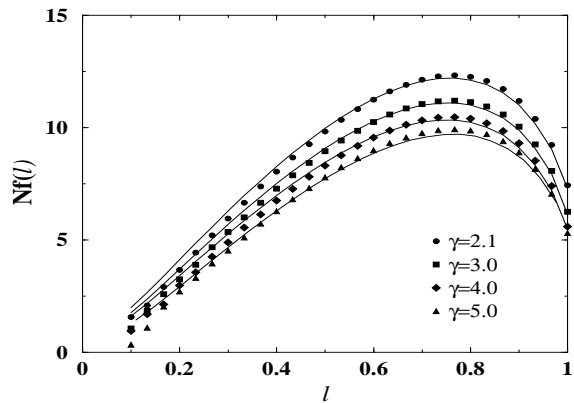}
\end{center}
\caption{\label{MR.fig}Number of loops for  MR networks without double
  links of $N=30$ nodes and different $\gamma$. The direct count
  averaged on $10$ realizations is compared to the analytic prediction
  for the same degree distribution (full lines).} 
\end{figure}

On the contrary, for the Molloy-Reed networks of same system size 
the direct count of loops is very close to the analytic
prediction. Fig. $\ref{MR.fig}$ reports the direct count of loops for
Molloy-Reed networks\cite{Nota} with $N=30$ and several degree distributions 
and it compares it with the corresponding analytic prediction.

\section{Conclusions}

In conclusion we have computed analytically the expected number of
loops of any size in a scale-free network.  We found that scale-free
graphs have a very large number of small loops compared to regular
random graphs.  On the contrary we have shown that, also with a
minimal connectivity $k_{min}\geq 3$ the expected number of Hamilton
cycles can be zero in the $N \rightarrow \infty $ limit provided that
$\gamma$ is sufficiently close to $2$. The reason for this is that
paths connecting many nodes need to pass frequently on nodes with high
connectivity. Put differently, it is not possible to embed a regular
graph of connectivity $c\ge 3$, which would have an Hamilton cycle, in
scale free networks if $\gamma$ is too small, even if all nodes have
$k_i\ge c$.  In the intermediate region of relatively large loops we
found that the expected number of loops attains its maximum for loops
of size $L\sim N$.  These results are derived both in the fitness and
in the Molloy-Reed ensembles. While the generic picture is the same,
the results in the two ensembles differ quantitatively highlighting
that the loop size distribution is somewhat sensitive to the precise
prescription for drawing random graphs. Moreover we have checked the
results with direct counting of computer generated scale-free networks
belonging to the two ensembles.  It would be desirable to derive
similar results for ensembles of correlated scale-free networks. 
\ack
G. B.  was partially supported by EVERGROW and by EU grant
HPRN-CT-2002-00319, STIPCO. We wish to thank F. Tria for useful
discussions.


\begin{thebibliography}{60}
\bibitem{Nota}Note that in the considered Molloy-Reed networks double links are not allowed. This correction is supposelly quite small for the considered system sizes.
\end{thebibliography}
\end{document}